\documentclass[preprint,showpacs,preprintnumbers,amsmath,amssymb]{revtex4}

\usepackage{graphicx}
\usepackage{dcolumn}
\usepackage{bm}
\usepackage{graphicx}
\usepackage{rotating}
\usepackage{dcolumn}
\usepackage{bm}

\begin{document}


\title{Evolution of Dipole-Type Blocking Life Cycles: Analytical Diagnoses and
Observations}
\author{ {Fei Huang$^{1, 2, 3}$}\thanks{Corresponding author address: Fei Huang, Department of
Marine Meteorology, Ocean University of China, 238 Songling Rd.,
Qingdao 266100, china. E-mail: huangf@mail.ouc.edu.cn}, Xiaoyan
Tang$^{1, 3}$, S. Y. Lou$^{1, 3}$ and Cuihua Lu$^4$\\
$^{1}$\small \it Department of Physics, Shanghai Jiao Tong
University,Shanghai,
200030, China\\
$^{2}$\small \it Physical Oceanography Laboratory, Department of
Marine Meteorology, \\Ocean\small \it  University of China,
Qingdao, 266003,
China\\
$^{3}$\small \it Center of Nonlinear Science, Ningbo University,
Ningbo,
315211, China\\
$^{4}$\small \it Weather Bureau of Zaozhuang, Zaozhuang, 277148,
China}
\begin{abstract}
A variable coefficient Korteweg de Vries (VCKdV) system is derived
by considering the time-dependent basic flow and boundary
conditions from a nonlinear, inviscid, nondissipative, and
equivalent barotropic vorticity equation in a beta-plane. One
analytical solution obtained from the VCKdV equation can be
successfully used to explain the evolution of atmospheric
dipole-type blocking (DB) life cycles. Analytical diagnoses show
that background mean westerlies have great influence on evolution
of DB during its life cycle. A weak westerly is necessary for
blocking development and the blocking life period shortens,
accompanied with the enhanced westerlies. The shear of the
background westerlies also plays an important role in the
evolution of blocking. The cyclonic shear is preferable for the
development of blocking but when the cyclonic shear increases, the
intensity of blocking decreases and the life period of DB becomes
shorter. Weak anticyclonic shear below a critical threshold is
also favorable for DB formation. Time-dependent background
westerly (TDW) in the life cycle of DB has some modulations on the
blocking life period and intensity due to the behavior of the mean
westerlies. Statistical analysis regarding the climatological
features of observed DB is also investigated. The Pacific is a
preferred region for DB, especially at high latitudes. These
features may associate with the weakest westerlies and the
particular westerly shear structure over the northwestern Pacific.
\end{abstract}
\pacs{47.32.-y; 47.35.-i; 92.60.-e; 02.30.Ik} \maketitle

\section{Introduction}
\label{usage.sec} Atmospheric blocking is an important large-scale
weather phenomena at mid-high latitudes in the atmosphere that has
a profound effect on local and regional climates in the immediate
blocking domain (Rex 1950a,b; Illari 1984) as well as in regions
upstream and/or downstream of the blocking event (Quiroz 1984;
White and Clark 1975). Therefore, it has long been of interest to
synoptic and dynamical meteorologists. The formation, maintenance
and collapse of atmospheric blocking always cause large-scale
weather or short-term climate anomalies. Hence, the prediction for
atmospheric blocking plays a significant role in regional midterm
weather forecast and short-term climate trend prediction. Because
of the needs of midterm weather forecast, there are has been a lot
of research on dynamics of blocking. However, the physical
mechanism leading to blocking formation remains unclear, which may
result in the difficulty of blocking onset prediction using
general circulation models (Tracton et al. 1989; Tibaldi and
Molteni 1990).

Theoretical (Long 1964; Charney and DeVore 1979; Tung and Linzen
1979; McWilliams 1980; Malguzzi and Malanotte-Rizzoli 1984; Haines
and Marshall 1987; Butchart et al. 1989; Michelangeli and Vautard
1998; Haines and Holland 1998; Luo 2000; Luo et al. 2001),
observational (Rex 1950a,b; Colucci and Alberta 1996; Berggren et
al. 1949; Illari 1984; Quiroz 1984; White and Clark 1975; Dole and
Gordon 1983; Dole 1986, 1989; Lejen\"and $\varnothing$land 1983;
Lupo and Smith 1995a), and numerical (Tanaka 1991, 1998; Luo et
al. 2002; D＊Andrea et al. 1998; Ji and Tibaldi 1983; Chen and
Juang 1992; Colucci and Baumhefner 1998) studies on the
atmospheric blocking systems have undergone substantial and
extensive development over the past several decades. Usually an
atmospheric blocking anticyclone has three type of patterns:
monopole-type blocking (or $\Omega$-type blocking), dipole-type
blocking (McWilliams 1980; Malguzzi and Malanotte-Rizzoli 1984;
Luo and Ji 1991), and multipole-type blocking (Berggren et al.
1949; Luo 2000). The setup of the two later patterns are usually
due to the strong nonlinear interaction between upstream high
frequency synoptic-scale transient eddies and planetary-scale
waves in exciting blocking circulation (Hansen and Chen 1982;
Shutts 1983, 1986; Ji and Tibaldi 1983; Illari and Marshall 1983;
Illari 1984; Hoskins et al. 1985; Egger et al. 1986; Mullen 1987;
Holopainen and Fortelius 1987; Haines and Marshall 1987; Vautard
and Legras 1988; Vautard et al. 1988; Chen and Juang 1992; Tanaka
1991; Lupo and Smith 1995b; Lupo 1997; Lupo and Smith 1998; Haines
and Holland 1998; Luo 2000, 2005a,b). However, the nature of the
planetary-synoptic scale interaction has yet to be clarified
theoretically (Colucci 1985, 1987).

Atmospheric blocking events, mainly located in the mid-high
latitudes and usually over the ocean, were first discovered as a
dipole pattern by Rex (1950a,b). Later, Malguzzi and
Malanotte-Rizzoli (1984) first found the dipole-type blocking from
the Korteweg-de Vries (KdV) Rossby soliton theory. However, their
analytical results cannot explain the onset, developing and decay
of a blocking system. Recently, Luo et al. (2001) proposed the
envelope Rossby soliton theory based on the deduced nonlinear
Schr\"odinger (NLS) type equations to explain the blocking life
cycle. However, they only obtained their solutions of the NLS type
equations numerically instead of analytically.

In 1895, the KdV equation was firstly derived in the classic paper
of Korteweg and de Vries (1895) as a fundamental equation
governing propagation of waves in shallow water. After that, much
progress has been made on this equation in both mathematics and
physics. Now, physical applications of the KdV equation have been
seen in a number of problems such as plasmas (Zabusky and Kruskal
1965; Rao et al. 1990), quasi-one-dimensional solid-state physics
(Flytzanis et al. 1985), nonlinear transmission lines (Yoshinaga
and Kakutani 1984), and so on.

In all these applications, the KdV equation arises as an
approximate equation valid in a certain asymptotic sense. Taking
account of additional physical factors, one may also obtain
various KdV-type equations. For instance, a variable coefficient
KdV-type (VCKdV) equation with conditions given at the inflow site
was derived as governing the spatial dynamics in a simplified
one-dimensional model for pulse wave propagation through
fluid-filled tubes with elastic walls (Flytzanis et al. 1985),
nonlinear transmission lines (Yoshinaga and Kakutani 1984), and so
on. From the discussion below of a problem on the dynamics of
atmospheric blockings, it will be seen that VCKdV equation is
fairly universal and useful to explain the lifetime of the
atmospheric blocking systems.

One important fact is that in the usual treatment of deriving
either KdV- or NLS-type equations in the study of atmospheric
blocking dynamics, one always separates the background flow,
assumed to linearly depend on $y$ only, and takes zero boundary
values. Nonetheless, in reality, the background westerly changes
with time as well as the shear type, which is something that has
not yet been considered. Therefore, in this paper, we are
motivated to introduce time into the background flow and boundary
conditions to reinvestigate the Rossby KdV theory for atmospheric
blocking systems.

The term ``blocking" denotes a breakdown in the prevailing
tropospheric westerly flow at midlatitudes, often associated with
a split in the zonal jet and with persistent ridging at higher
latitudes (Rex 1950a,b; Illari 1984). Therefore, the basic pattern
of blocking is dipole type and it always occurs under weak
background westerlies (Shutts 1983; Luo and Ji 1991; Luo 1994; Luo
et al. 2001; Luo 2005b). Weak background westerlies as a
precondition for block onset were first noted by Shutts (1983),
who found that the block does not occur for more rapid flows in
which a stationary free state cannot be excited. However, Tsou and
Smith (1990) and Colucci and Alberta (1996) suggested that the
preconditioned planetary-scale ridge (incipient blocking ridge)
may be another necessary condition for block onset in addition to
the weak planetary-scale westerly flow.

 Although previous
theoretical analysis have shown that the weaker background
westerlies were prevailing for the onset of DB than that of
$\Omega$-type blocking (Luo 1994), the role of the weak westerlies
on the evolution of DB life cycle is still unclear. Further, how
variations of the time-depend background westerlies influence the
evolution of DB during its life period is also an interesting
problem, since the realtime westerlies vary with time. Shear in
the background westerlies was always introduced into theoretical
models (Tung and Linzen 1979; McWilliams 1980; Malguzzi and
Malanotte-Rizzoli 1984; Haines and Marshall 1987; Butchart et al.
1989; Luo 1994; Haines and Holland 1998; Luo et al. 2001) in the
study of blocking, while Luo (1995) revealed that dipole structure
of solitary Rossby waves excited by the $\beta$ parameter could be
obtained excluding the effect of shearing basic-state flow. In
addition, the role of anticyclonic shear of basic-state flow on DB
was emphasized (Luo 1994). In contrast, Luo et al. (2001) found
that the cyclonic shear of the background westerly wind at the
beginning of block onset was a favorable preblock environment,
which increased the strength of the precursor blocking ridge, but
the anticyclonic shear weakened the precursor blocking ridge
considerably. Recently, Luo (2005b) also found that the asymmetry,
intensity, and persistence of dipole block depend strongly upon
the horizontal shear of the basicstate flow prior to block onset.
Then there arises another question: what does the role of
cyclonic/anticyclonic shear of the background westerlies play in
the evolution of DB during different stages of its life cycle?

In this paper, we will focus on the two questions mentioned above
by diagnosing the analytical solution obtained from a VCKdV
equation derived from the atmospheric nonlinear, inviscid,
nondissipative, and equivalent barotropic vorticity equation in a
beta plane considering the time-dependent background flow and
boundary conditions. The paper is organized as follows. Section 2
introduces the derivation of the VCKdV equation and its analytical
solution. Analytical diagnosis of background westerly variations
(including the variation of mean flow, westerly shears, and the
time-dependent background flow) on the evolution of DB during its
life cycle are investigated in Section 3. In the following
section, some statistical observational results about DB in the
Northern Hemisphere are given and possible interpretations from
above theoretical model are demonstrated. Section 5 will outline
our conclusions.

\section{Theoretical model}
\subsection{Derivation of the VCKdV equation}

Our starting model is the atmospheric nonlinear, inviscid,
nondissipative, and equivalent barotropic vorticity equation in a
beta-plane channel (Pedlosky 1979; Luo 2001, 2005a):
\begin{equation}\label{ibnv0}
\left(\partial_t+u\partial_x+v\partial_y\right)(v_x-u_y-F\psi)+\beta\psi_x=0,
\end{equation}
which is well known and one of the most important models in the
study of atmospheric and oceanic dynamical systems. In equation
(\ref{ibnv0}), $u=-\psi_y,~v=\psi_x $, $\psi$ is the dimensionless
stream function, $u$ and $v$ are components of the dimensionless
velocity, $F=L^2/R_0^2$ is the square of the ratio of the
characteristic horizontal length scale $L$ to the Rossby
deformation radius $R_0$, $\beta=\beta_0(L^2/U)$,
$\beta_0=(2\omega_0/a_0) \cos\phi_0$, in which $a_0$ is the
Earth's radius, $\omega_0$ is the angular frequency of the earth's
rotation, $\phi_0$ is the latitude, and $U$ is the characteristic
velocity scale. The characteristic horizontal length scale can be
$L=10^6${\rm m} and the characteristic horizontal velocity scale
is taken as $U=1 ms^{-1}$.

Similar to the usual treatments, we rewrite the stream function as
$\psi=\psi_0(y,\ t)+\psi'$ where $\psi_0$ means the background
flow term, and introduce the stretched variables
$\xi=\epsilon(x-c_0t),~\tau=\epsilon^3t$, ($c_0$ is a constant).
In the previous studies, the background field $\psi_0$ is often
taken only as a linear function of $y$, and in most cases one
simply makes the expansion
$\psi'=\sum_{n=1}^{\infty}\epsilon^n\psi'_n(\xi,y,\tau)$. However,
here we also have the expansion for the background flow term as
$\psi_0=U_0(y)+\sum_{n=1}^{\infty}\epsilon^n U_n(y,\tau)$. Then,
again similarly, substitute the expansions into (\ref{ibnv0}) and
set the coefficient of each order of $\epsilon$ equal to zero. For
notation simplicity, the primes are dropped out in the remaining
of this paper. In the first order of $\epsilon$, we obtain

$\psi_1=A(\xi,\tau)G(y,\tau)$, \\where $G$ is determined by\\
$(U_{0y}+c_0)G_{yy}-(F^2c_0+\beta+U_{0yyy})G=0.$ \\In the second
order, we have\\
$\psi_2=A(\xi,\tau)G_1(y,\tau)+A^2(\xi,\tau)G_2(y,\tau)$, \\where
$G_1$ satisfies\\
$(U_{0y}+c_0)^2G_{1yy}-(U_{0y}+c_0)(F^2c_0+U_{0yyy}+\beta)G_1
-[(U_{0y}+c_0)U_{1yyy}-(F^2c_0+U_{0yyy}+\beta)U_{1y}]G=0,$ \\and
$G_2$ satisfies\\
$2(U_{0y}+c_0)^2[(U_{0y}+c_0)G_{2yy}-(F^2c_0+U_{0yyy}+\beta)G_{2}]+[(F^2c_0+U_{0yyy}
+\beta)U_{0yy}\\-(U_{0y}+c_0)U_{0yyyy}]G^2=0.$

Substituting the solutions obtained from the previous two orders
and $\psi_3=0$ into the coefficient of the third order of
$\epsilon$, then integrating the result with respect to $y$ from 0
to $y_0$ as in the usual treatments of the oceanic and atmospheric
dynamics, yield the modified VCKdV equation ($e_2\neq 0$) and/or
VCKdV equation ($e_2=0$)
\begin{eqnarray}
e_1A_{\xi\xi\xi}+(e_2A^2+e_3A+e_4)A_{\xi}+(e_{5}A)_\tau+e_6=0,
\end{eqnarray}
with ($e_i\equiv
e_i(\tau))$,\\$~e_1=\int_0^{y_0}-G(U_{0y}+c_0){\rm d}y$, \\$~
e_2=\int_0^{y_0}2(G_{yyy}G_2-G_yG_{2yy})+GG_{2yyy}
-G_{yy}G_{2y}{\rm d}y$, \\$~
e_3=\int_0^{y_0}2U_{1yyy}G_2-2U_{1y}G_{2yy}+G_{yyy}G_1-(G_{y}G_{1y})_y+GG_{1yyy}{\rm
d}y$, \\$~
e_4=\int_0^{y_0}U_{1yyy}G_1-G_{yy}U_{2y}+GU_{2yyy}-U_{1y}G_{1yy}{\rm
d}y,$\\$~ e_5=\int_0^{y_0}G_{yy}-F^2G{\rm d}y,$ \\$~
e_6=\int_0^{y_0}(U_{1yy}-F^2U_{1})_{\tau}{\rm d}y$.

\subsection{Exact solution of the VCKdV equation}

Obviously, in the above derivation of the VCKdV equation, the
background flow is left as an arbitrary function of $\{y,t\}$. For
the sake of obtaining meaningful analytical solutions to the VCKdV
equation, we have to fix its variable coefficients further. In the
expansion of the background flow term $\psi_0$, if $U_n=0$ for
$n\geq 1$, it becomes a function of $y$ as usual. Therefore, the
higher order items $U_n$ can be viewed as higher-order corrections
to the lowest one, $U_0$. Because of this, we also take $U_0$ as a
linear function of $y$; that is, $U_0=a_1y+a_0$ with $a_0,~a_1$
constant, and for the next higher order, as mentioned before, it
is chosen as a quadratic function of $y$ with time dependent
coefficients; that is, $U_1=a_2y^2+a_3y+a_4$ with $a_2,~a_3,~a_4$
all time dependent functions. The higher orders are all taken as
zero, namely $U_n=0$ for $n\geq 2$ so as to simplify the problem.
In this case, the background flow is similar to the cases in
Gottwald and Grimshaw (1999) except that the mean flow
configurations are time-varying instead of stable. It is noted
that the quantity $\epsilon a_2$ plays the role of the shear of
the background flow, and the background westerly wind is
determined by $a_1+\epsilon a_3$. Apparently, both the shear and
the background westerlies vary with time. After these selections,
$G,~G_1$ and $G_2$ can be easily solved as\\ $G=F_1\sin(My+F_2)$,
\\$~G_1=F_5\sin(My+F_6)+\frac {(F^2+M^2)F_1} {4M(a_1F^2-\beta)}
[M(2a_2y+a_3)\sin(My+F_2)-(2a_2M^2y^2+2a_3M^2y-a_2)\cos(My+F_2)]$,
\\$G_2=F_3\sin(My+F_4)$, \\where $c_0=-\frac {a_1M^2+\beta}
{F^2+M^2}$. $F_1\sim F_6$, and all functions with respect to time
should be determined by the boundary conditions. No constraints
are made on these functions because in the following, we will
focus on the behavior of the VCKdV equation with different values
for the parameters appeared in the background flow so as to
investigate what effect of the background can make on the possible
coherent structure; that is, the soliton solution.

With the above selections and results, we can finally write down
an exact solution to the VCKdV equation as
\begin{eqnarray}
&A=&\frac {C_2\xi-H} {2C_3e_6\sqrt{-C_1+C_2\tau}}+\frac
{(F^2a_1-\beta)^{1/3}} {e_6(F^2+M^2)^{2/3}\sqrt{-C_1+C_2\tau}}
\left\{\frac {C_4} {C_3^2C_7^{1/3}} \right.\nonumber\\&&\left.
+12C_3K^2C_7^{2/3}{\rm sech}^2\left[\frac
{8K^3C_3^3C_7-2KC_4}{C_2\sqrt{-C_1+C_2\tau}}
+KC_5-4K^3C_6\right.\right.\nonumber\\&&\left.\left.+\frac
{KC_3C_7^{1/3}(F^2+M^2)^{2/3}} {2(F^2a_1-\beta)^{1/3}}\left(\frac
{2\xi}{\sqrt{-C_1+C_2\tau}}\right.\right.\right.\nonumber\\&&\left.\left.\left.+\int\frac
H{(-C_1+C_2\tau)^{3/2}}{\rm d}\tau\right)\right]\right\}-\frac
{e_4\sqrt{-C_1+C_2\tau}}{C_3e_6^2},
\end{eqnarray}
where $H=\int\frac
{2(C_1-C_2\tau)(e_4e_{6\tau}-e_{4\tau}e_6)}{e_6^2}
+2e_7C_3\sqrt{C_2-C_1\tau}-\frac {e_4C_2} {e_6} {\rm d }\tau$,
\\$~
e_4=[2a_2\sin(My_0+F_6)-2a_2\sin(F_6)-M(2a_2y_0+a_3)\cos(My_0+F_6)+Ma_3\cos(F_6)]F_5-\frac
{(F^2+M^2)F_1}
{4(F^2a_1-\beta)M}[a_2a_3\sin(F_2)M-(M^2a_3^2+2a_2^2)(\cos(My_0+F_2)-\cos(F_2))
+M(2a_2y_0+a_3)(2M^2a_2y_0^2+2M^2a_3y_0-a_2)\sin(My_0+F_2))]$,\\$
~ e_6=-\frac {F_1(M^2+F^2)} M[\cos(F_2)-\cos(My_0+F_2)]$,\\$ ~
e_7=-\frac {y_0}
6[(2y_0^2F^2-12)a_{2\tau}+3y_0F^2a_{3\tau}+6F^2a_{4\tau}]$, \\and
$M, K, C_i,(i=1,2,...,7)$ are arbitrary constants. There is also a
necessary relation between $F_2\sim F_4$: \\
$F_3[2M^2(\lambda0^2a_1-\beta)\sqrt{-C_1+C_2\tau}
(2a_2\sin(F_4)-2a_2\sin(My_0+F_4)-a_3M\cos(F_4)+M(2y_0a_2+a_3)\cos(My_0+F_4))]-F_1^2(F^2+M^2)
[a_2M^3\sqrt{-C_1+C_2\tau}(\cos(F_2)\sin(F_2)
-\cos(My_0+F_2)\sin(My_0+F_2)+My_0)-C_3(F^2a_1-\beta)(F^2+M^2)(\cos(F_2)-\cos(My_0+F_2))^2]=0$.

\section{Analytical diagnosis on DB evolution}
\label{theory}

In this section we analyze a DB evolution case under certain
parameters. Set $e_4=0$ and $a_4=-\frac {y_0} 2a_3+(\frac 2
{F^2}-\frac {y_0^2} 3)a_2$ (and then $e_6=0$).
 Hence, one possible approximate solution to
(\ref{ibnv0}) is in the form of
\begin{eqnarray}
&\psi \approx&\epsilon a_2y^2+(a_1+\epsilon a_3)y+a_0-\frac {y_0}
2\epsilon a_3+\left(\frac 2 {F^2}-\frac {y_0^2} 3\right)\epsilon
a_2\nonumber\\&&-\frac {\sin(My+F_2)}{[\cos(F_2)-\cos(My_0+F_2)]}
\quad\times\frac {\epsilon M} {(M^2+F^2)
\sqrt{-C_1+\epsilon^3C_2t}}\left\{\frac {\epsilon C_2}
{2C_3}(x-c_0t)\right.\nonumber\\&&\left.+\frac {C_4M_1}
{C_3^2C_7^{1/3}}+12C_3M_1K^2C_7^{2/3} \times{\rm
sech}^2\left[\frac {K} {\sqrt{-C_1+\epsilon^3C_2t}}\left(\frac
{\epsilon C_3C_7^{1/3}}
{M_1}(x-c_0t)\right.\right.\right.\nonumber\\&&\left.\left.\left.+\frac
{8K^2C_3^3C_7-2C_4}
{C_2}\right)+K(C_5-4K^2C_6)\right]\right\},\label{rol1}
\end{eqnarray}
with $M_1=\frac {(F^2a_1-\beta)^{1/3}} {(F^2+M^2)^{2/3}}$.

From the analytical solution of streamfunction $\psi$ it is easily
to derive the background westerly flow
$\overline{u}=-\frac{\partial \psi_0}{\partial
y}=\overline{u}_0+\delta y, $ where $\overline{u}_0=-(a_1+\epsilon
a_3(t))$ and $\delta=-2\epsilon a_2(t)$. As we known, the weather
or climate periodically changes on either large or small scales.
Thus, it is reasonable to choose the time-dependent unknown
functions in the background flow as periodic, so the sine function
is in use. Thus we choose $a_2(t)=-a_{20}sin(k_2t),
a_3(t)=-a_{30}sin(k_3t)$. The basic-state westerlies have cyclonic
shear when $\delta<0$ or $a_{20}>0$, but correspond to
anticyclonic westerly shear when $\delta>0$ or $a_{20}<0$. Here we
consider $\phi_0=60^{o}N$ in that the blocking systems are mainly
situated in the mid-high latitudes. For example, when we take the
functions and parameters in our analytical solution (\ref{rol1})
as $a_{20}=0.1, k_2=2.1, a_{30}=0.5, k_3=k_2, ~
F_2=\arccos(10^{-8}{\rm cosh}{(0.75t-4.5)}),~\epsilon =0.1,~ a_0 =
2,~a_1=-0.2,~C_1 = -30,~ C_2 = 0.1,~ C_3 = -10,~ C_4=-16, ~C_5 =
4,~ C_6 = -7,~ C_7 = 0.1,~ K = 0.2,~F=64,~ M=\frac {\pi} {6}$, a
dipole-type blocking life cycle appears and is shown in Fig. 1.

\begin{figure}
\includegraphics[width=25pc,height=13pc,angle=0]{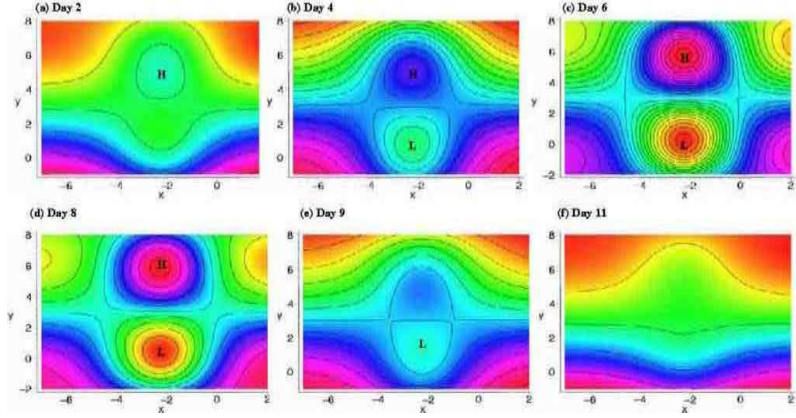}
\centering\caption{A dipole blocking life cycle from the
theoretical solution (\ref{rol1}) with $a_2=-0.1\sin(2.1t),
a_3=-0.5\sin(2.1t), \epsilon=0.1, a_1=-0.2$. The contour interval
CI=0.2.}
\end{figure}

Figure 1 clearly reveals the onset, development, maintenance, and
decay processes--a whole life cycle of a dipole-type blocking
event. The streamlines are gradually deformed and split into two
branches at the second day, with the anticyclonic high in the
north developing first (Fig. 1a). Then the cut-off low develops
south of the high, forming a pair of high-low dipole pattern where
the center of the high is located at the higher latitude than that
of the low (Fig. 1b). They are strengthened daily. At around the
sixth day (Fig. 1c), they are at their strongest stage and then
become weaker and eventually disappear after the eleventh day
(Fig. 1d-f). Obviously, Fig. 1 possesses the phenomenon's salient
features including their spatial-scale and structure, amplitude,
life cycle and duration. Therefore, Fig. 1 is a very typical
dipole-type blocking episode. More importantly, it corresponds
quite well to a real observational blocking case (Fig. 2) that
happened over the Pacific during 2-12 January 1996, which is
obtained from the National Centers for Environmental
Prediction-National Center for Atmospheric Research (NCEP-NCAR)
reanalysis data. Because blocking is a large-scale atmospheric
phenomenon, the geopotential height fields in Fig. 2 were filtered
by preserving wave numbers 0-4 in a harmonic analysis around
latitudes in order to filter high-frequency synoptic-scale
perturbations.

\begin{figure}
\includegraphics[width=20pc,angle=90]{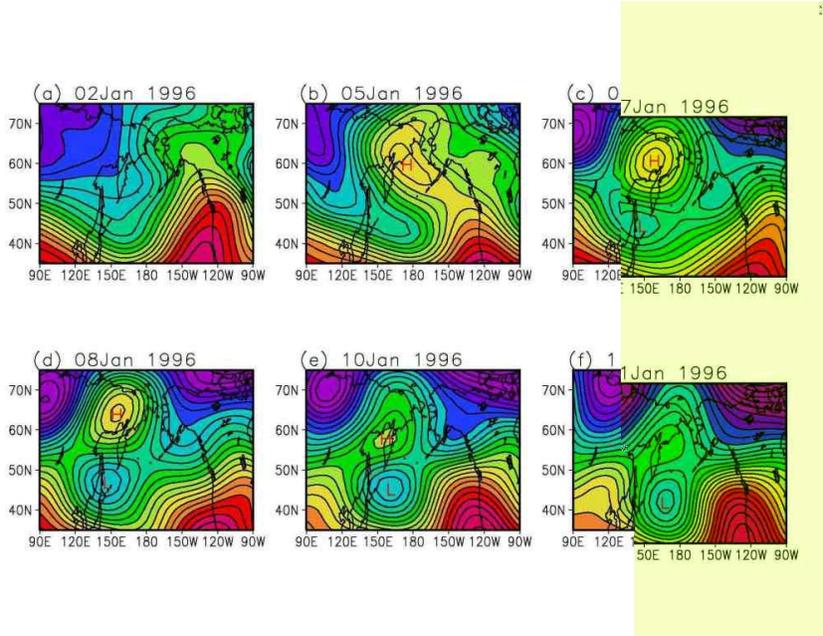}
\centering\caption{Filtered geopotential height at 500-hPa
pressure level of a blocking case during 2-11 Jan 1996. Contour
interval is $4$ gpdm. The $x$-axis is longitude, and the $y$-axis
is latitude.}
\end{figure}

It is easily found in Fig. 2 that the life cycle of this blocking
lasts almost 10 days, experiencing three stages: onset (2-5
January, 1996), maturity (7-9 January, 1996) and decay (10-12
January, 1996). This blocking event developed within an
atmospheric long Rossby wave ridge extending northwestward and an
upstream cutoff low moving southeastward (Fig. 2a, b). With the
strengthening of the blocking high and the cutoff low, two closed
centers of high and low appeared forming a north-south dipole-like
pattern with the closed high in the north and the closed low in
the south during the mature stage (Fig. 2c, d). Then the high
center decayed and vanished first, as well as the cutoff low
decayed subsequently (Fig. 2e,f). At last the block became a
travelling Rossby wave ridge and moved eastward (figure omitted).

\subsection{Effect of the mean background westerlies}
Previous research showed that the background westerlies are a
necessary precondition for the onset of anticyclonic blocking and
they also played an important role in the blocking life cycle
(Shutts 1983, 1986; Luo and Ji 1991; Luo 1994; Luo et al. 2001).
For DB evolution during its life cycle, the role of the mean
background westerlies in the DB evolution is not clear in a VCKdV
soliton system, while the typical KdV Rossby soliton can only
represent a steady-state DB pattern instead of an evolution of DB
life cycles. To simplify the question, here only variation of the
mean flow is considered without westerly shear ($a_2(t)=0$) and
time-dependent background westerly term ($a_3(t)=0$). Only the
parameter $a_1$ varies in different constants indicating the
variation of the mean background westerlies. The evolutions of DB
life cycles under different mean background westerlies are shown
in Fig. 3.

\input epsf
\begin{figure}
     \centering\epsfxsize=13cm\epsfysize=16cm\epsfbox{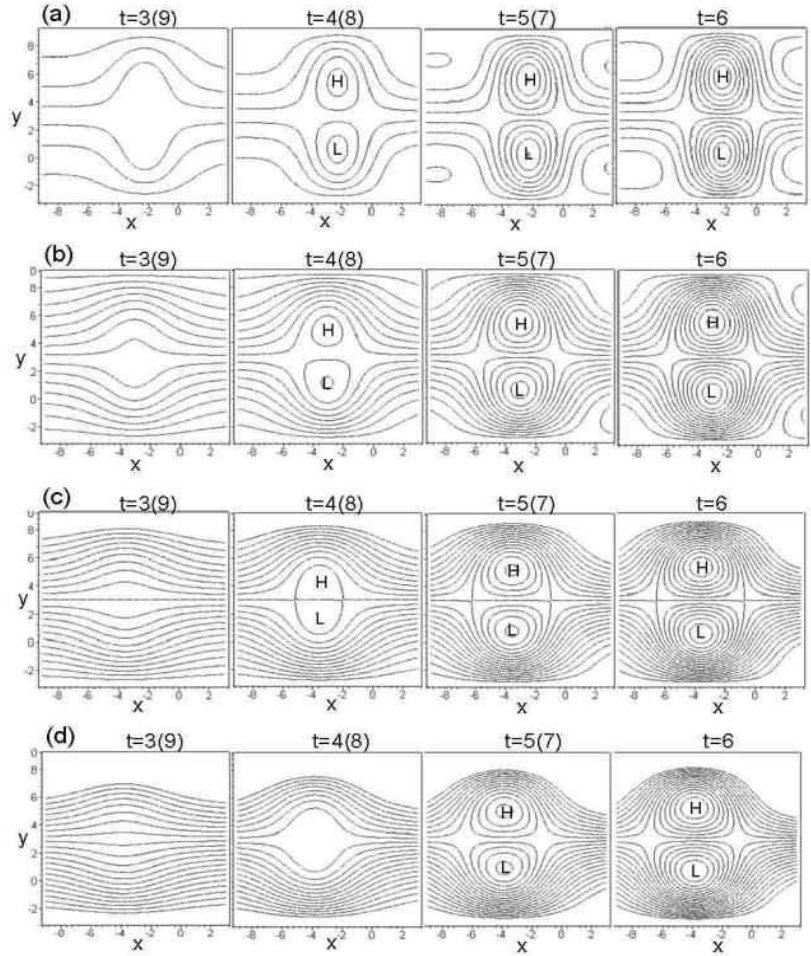}
\caption{The evolutions of DB life cycles under different
background westerlies (a) $a_1=-0.2$; (b) $a_1=-0.5$; (c)
$a_1=-0.8$; (d) $a_1=-1.0$. CI=0.4.}
\end{figure}

It is shown in Fig. 3 that the intensity, zonal and meridional
scales, position, and period of DB during its life episode alter
with the increasing mean background westerlies. Under the weak
mean westerly $a_1=-0.2$ condition (Fig. 3a), the DB period is
about 10 days from $t=1$ to $t=10$, consistent with typical period
of blocking observed \citep{rexb, White, Lejenas, Dole89,
Luodipole, Lupo95, huangphd}. At time $t=2$ an incipient
dipole-like envelop Rossby soliton pattern appears and begins to
enhance, forming closed high and low centers at $t=4$. Then the
high and low centers strengthen dramatically from $t=4$ to $t=6$
and reach their peaks at $t=6$. Subsequently, the high and low
centers in DB decrease gradually from $t=7$ to $t=10$. The
meridional scale of DB (distance between the high and low centers)
increases during the developing period ($t=1$ to $t=6$) of DB life
cycle and decreases during the decay process ($t=7$ to $t=10$). It
is noticeable that the high and low centers in the DB streamlines
pattern seem antisymmetric along the north-south direction as well
as the evolution process is symmetric with respect to the mature
phase of DB at $t=6$ under the high idealized theoretical
solution. However, the main characters of the DB streamlines
pattern and its evolution, including the onset, development,
mature, and decay processes, are well captured.

Comparing the DB evolution processes under different mean
westerlies (Fig. 3b-d), it is easily found that the incipient
dipole-like envelop at time $t=3$ weakens with the mean westerly
increasing. The closed streamlines of high and low centers in DB
under weak westerly conditions (Fig. 3a,b) at $t=4$ disappear and
are replaced by the gradually decreasing envelop streamline
pattern (Fig. 3c,d) when the parameter $a_1$ varies from
$a_1=-0.8$ to $a_1=-1.0$. That means the period of the DB life
cycle shortens when the mean westerlies increas. At time $t=4$ or
$t=8$, it is very clear that the DB intensity represented by
values of the DB high and low centers also weakens with the mean
flow increasing. This feature is also clearly seen in Fig. 4,
showing the intensity of DB high (Fig. 4a) and low (Fig. 4b)
center varying with respect to time $t$. It is obvious that the
high center of DB weakens while the low center strengthens when
the mean westerlies increas throughout the period of DB life
cycles. Nevertheless, the behaviors of the high and low centers
appear inconsistent during the different developing stages of DB
life episode. That is, the high center (Fig. 4a) decreases greatly
during the onset ($t=1$ to $t=4$) and decay ($t=8$ to $t=11$)
stages of DB life episode, and decreases slowly in the DB mature
phase ($t=5$ to $t=7$). On the contrary, the low center (Fig. 4b)
deepens greatly during the mature stage of DB life cycle, but
strengthens relatively slightly at the onset and decay stages.

\begin{figure}
\includegraphics[bb=100 300 600 600, scale=0.95]{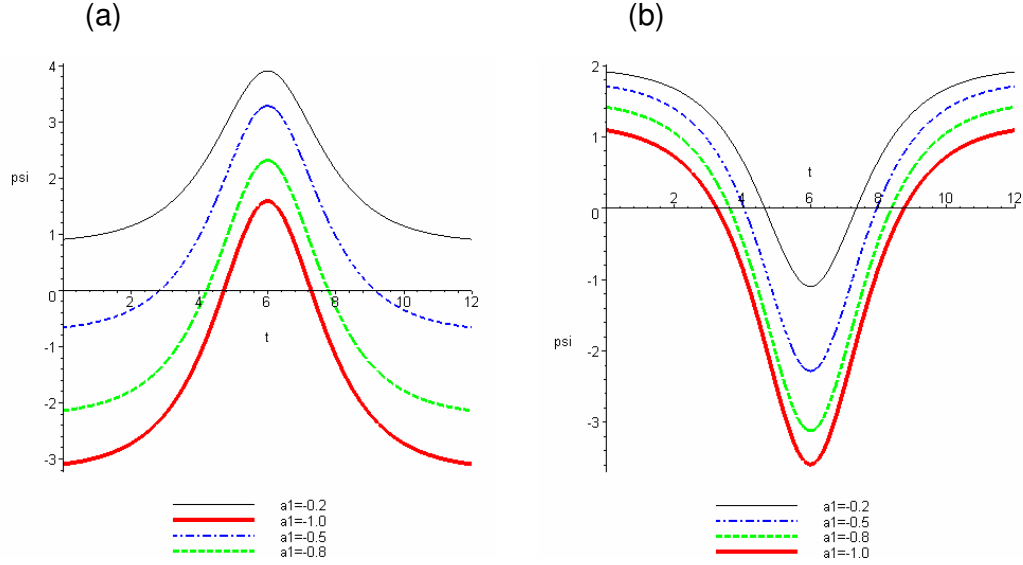}
\centering\caption{The intensity of the DB high (a) and low (b)
center varying with respect to time $t$.}
\end{figure}

To see more clearly the movement of DB high center with respect to
time $t$, Fig. 5 shows space-time ($x-t$ or $y-t$) cross sections
crossing the high center. From the $x-t$ cross section (Fig. 5a)
it is found that the blocking high moves westward and the zonal
scale of DB enlarges with the strengthening of the mean
westerlies. This feature can also be identified in Fig. 3. For the
$y-t$ cross section crossing the high center (Fig. 5b), the
meridional scale of the DB shortens and the evolution crossing the
dipole high/low center with respect to $t$ displays a shrinking
dipole-like pattern, suggesting shortening of DB period of life
cycles under the condition of the mean westerlies increasing. This
result implies that the DB could not maintain stationary long time
in strong westerlies (Shutts 1983).

\input epsf
\begin{figure}
     \centering\epsfxsize=12cm\epsfysize=14cm\epsfbox{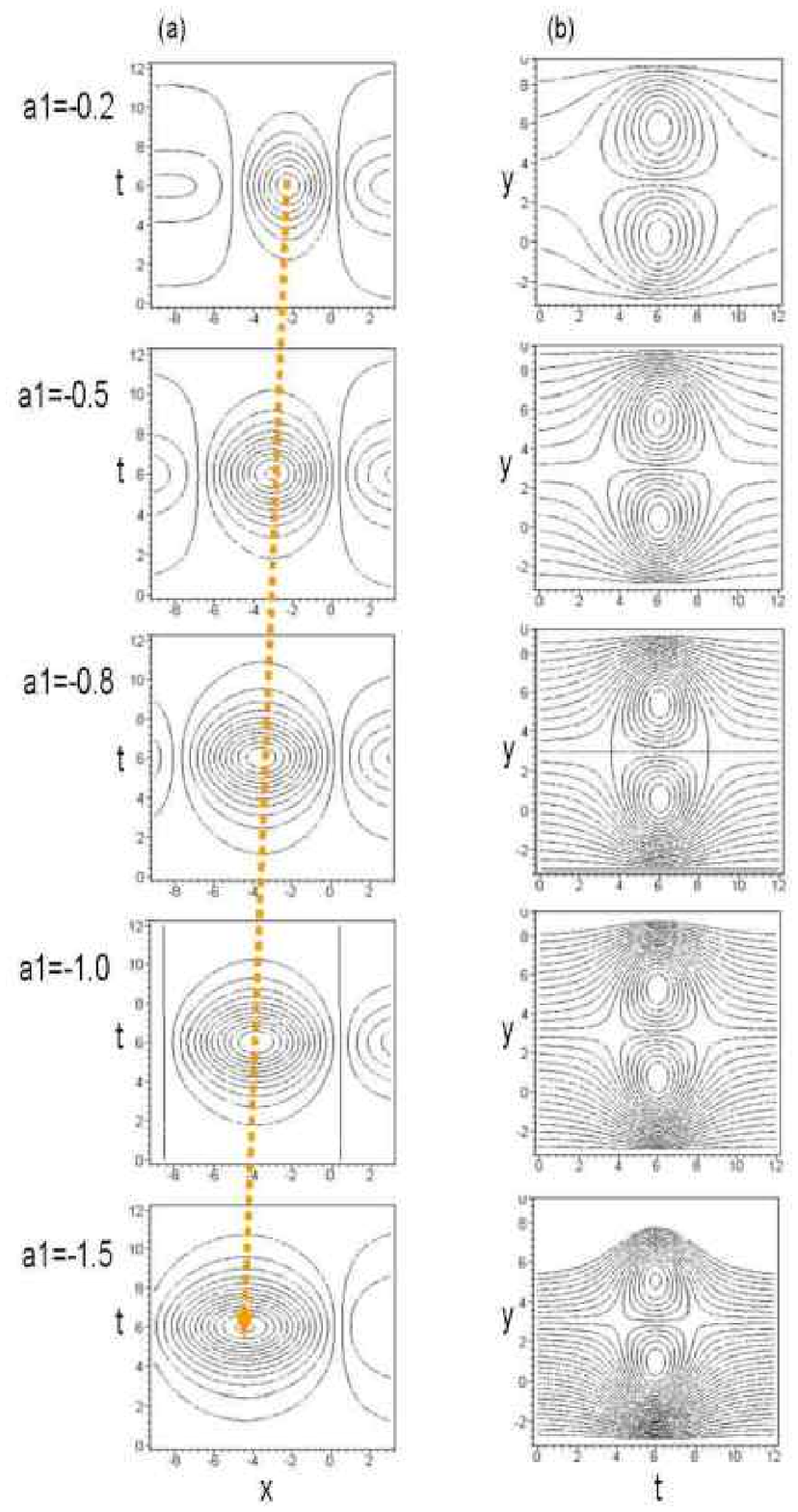}
\centering\caption{The space-time $x-t$ (a) and $y-t$ (b) cross
sections crossing the high center of DB under different basic mean
westerlies. CI=0.2.}
\end{figure}

\subsection{Effect of the background westerly shear}

In this section effects of the westerly shears including cyclonic
shear and anticyclonic shear on evolution of DB are investigated.
As mentioned above, the westerly shear parameter $\delta$ is
expressed as $\delta=-2\epsilon a_2(t)$; $\delta<0$ or $a_2(t)>0$
denotes the cyclonic background westerly shear, and $\delta>0$ or
$a_2(t)<0$ represents the anticyclonic shear. As we know, $a_2(t)$
is a function with respect to time $t$. To simplify the question,
here $a_2$ is assumed to be a small constant, suggesting a
time-independent weak linear shear superposed on the background
mean flow. The mean westerlies without the time-dependent term
($a_3(t)=0$) is also assumed. Consequently, the background
westerly with linear shear is $\overline{u}=\overline{u}_0+\delta
y$, where the mean flow $\overline{u}_0=-a_1=0.6$.

\subsubsection{The cyclonic westerly shear}

The background westerly shear is always introduced into a
theoretical model (Tung and Linzen 1979; McWilliams 1980; Malguzzi
and Malanotte-Rizzoli 1984; Haines and Marshall 1987; Butchart et
al. 1989; Luo 1994; Haines and Holland 1998; Luo et al. 2001; Luo
2005b) in the study of blocking, but the role of the cyclonic
westerly shear (CWS) in evolution of DB during its life cycle is
not clear, although Luo et al. (2001) and Luo (2005b) emphasized
the important role of the cyclonic shear of background westerly
wind in block onset. Recently, Dong and Colucci (2005) also
demonstrated the important effect of cyclonically sheared flow on
forcing weakening westerlies associated with the Southern
Hemisphere blocking onset. To compare the influence of different
cyclonic shear on a DB episode, the DB evolutions at parameter
$a_2=0.03, 0.15, 0.3, 0.45$ and $0.6$, corresponding to 1\%, 5\%,
10\%, 15\% and 20\% of the mean westerlies $\overline{u}_0$, are
investigated respectively in Fig. 6. Results show that weak CWS is
favorable for the onset of DB, similar to the conclusion by Luo et
al. (2001), Luo (2005b), and Dong and Colucci (2005). However, the
DB life period shortens and the DB high/low centers weaken
simultaneously when the weak CWS enhances during the evolution of
DB episode (Figs. 7a, c). The intensity variation of the low
center appears different behaviors, considering the CWS
comparisons with and without shear (Fig. 4). That is, the low
weakens along with increasing CWS (Fig. 7c), and strengthens with
increasing background mean westerlies without shear (Fig. 4b). A
noticeable phenomenon is that the CWS destroys the antisymmetric
structure of the DB high/low poles. The low develops more slowly
than the high at the DB onset stage ($t=4$), while it trails off
faster than the high during the decay period ($t=8$) with the CWS
increasing (Fig. 6). This feature is also indicated in Figs. 7a,c.

\input epsf
\begin{figure}
     \centering\epsfxsize=13cm\epsfysize=16cm\epsfbox{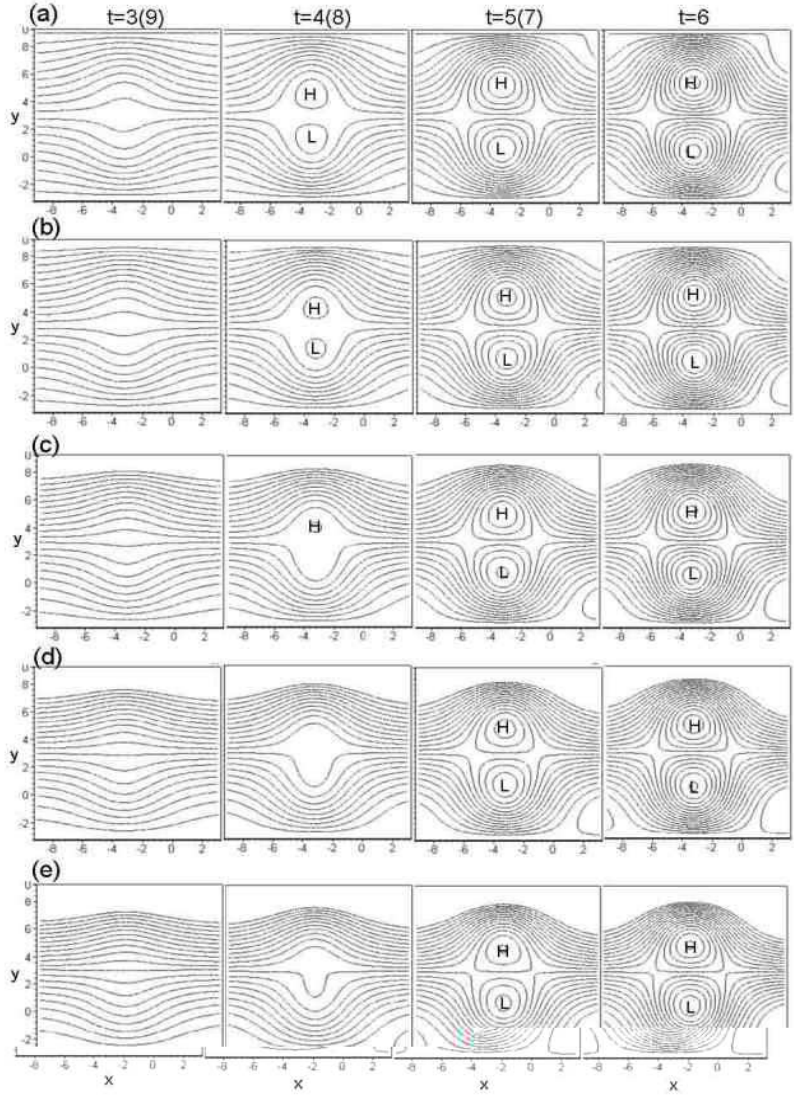}
\centering\caption{The stream function patterns of DB evolution
under different cyclonic westerly shears. (a) $a_2=0.03$; (b)
$a_2=0.15$; (c) $a_2=0.3$; (d) $a_2=0.45$; (e) $a_2=0.6$. CI=0.4.}
\end{figure}

\begin{figure}
\includegraphics[bb=100 200 600 600,scale=0.95]{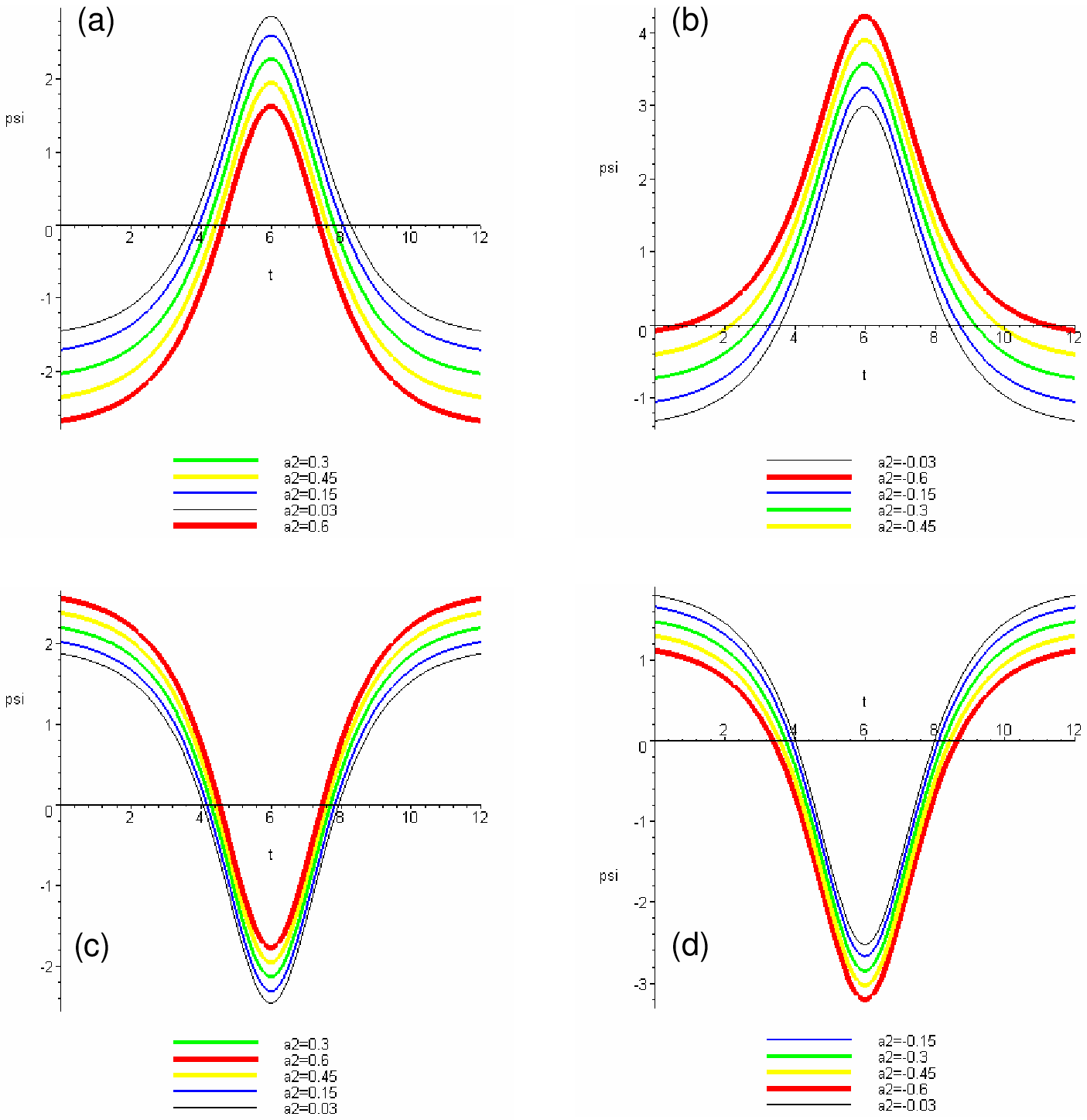}
\centering\caption{The intensity of the DB high (a, b) and low (c,
d) center varying with respect to time $t$ under different
conditions of westerly shears. (a) and (c) are for cyclonic shear,
(b) and (d) are for anticyclonic shear.}
\end{figure}

\subsubsection{The anticyclonic westerly shear}

Luo (1994) revealed that strong anticyclonic westerly shear (AWS)
was disadvantageous for establishment of dipole-type blocking, but
his solution could not interpret the onset and decay process of
DB. Recently, Luo et al. (2001) and Luo (2005b) also pointed out
that the anticyclonic background westerly shear weakened the
precursor blocking ridge considerably; therefore, the formation of
a blocking anticyclone was difficult. However, the role of the AWS
on DB life cycles is not clear. In our model, the influence of
different weak AWS on DB episode with parameters $a_2=-0.03,
-0.15, -0.3, -0.45$, and $-0.6$, respectively, are compared with
each other in Fig. 8. Results show that only very weak
anticyclonic shear is more preferable for the onset and
maintenance of DB, resulting longer life of DB (Figs. 8a-c). There
exists a threshold value of $a_2$ limiting the anticyclonic shear
that controls the appearance of DB streamline patterns. Under the
mean westerlies $\overline{u}_0=0.6$, the threshold of $a_2$ is
about $-0.45$, or the threshold shear $\delta_c=0.09$, denoting
the slope of lines on the ($\overline{u}_0-y$) plane. This means
the DB is easily established under weak AWS condition when
$\delta<\delta_c$. For $\delta<\delta_c$, the DB life period is
prolonged obviously and DB establishes (decays) earlier (later)
with the increasing of AWS (Figs. 8a-c). For $\delta>\delta_c$,
the dipole-like circulation can still develop, but never be a
blocking pattern splitting the westerlies into two northward and
southward branches (Figs. 8d,e). Instead, the separate northward
westerlies around the high disappear and the developed high in the
dipole comes from the most northern boundary, which indicates a
cutoff high from the North Pole region moving southward, forming
the dipole circulation together with the low in the south.
Actually, in a daily synoptic chart, the cutoff high from the
North Pole region is often seen moving southward. The threshold
value of AWS decreases along with decreasing of the mean
background westerlies $\overline{u}_0$.

It is also obvious that the AWS induces the asymmetry development
of the high/low centers in DB during different stages of its life
cycle (Fig. 8). Especially at DB onset ($t=3, 4$) and decay ($t=8,
9$) phases, the high center is apparently stronger than the low
center, and the high/low center tends to strengthen along with the
strengthening of the AWS. The DB high/low center strengthens
simultaneously when the AWS strengthens during the evolution of DB
episode (Figs. 7b, d), which appears to have opposite behavior
than that with CWS (Figs. 7a, c). In addition, the intensity of
the DB dipole centers with AWS is stronger than that of CWS.

\input epsf
\begin{figure}
     \centering\epsfxsize=13cm\epsfysize=16cm\epsfbox{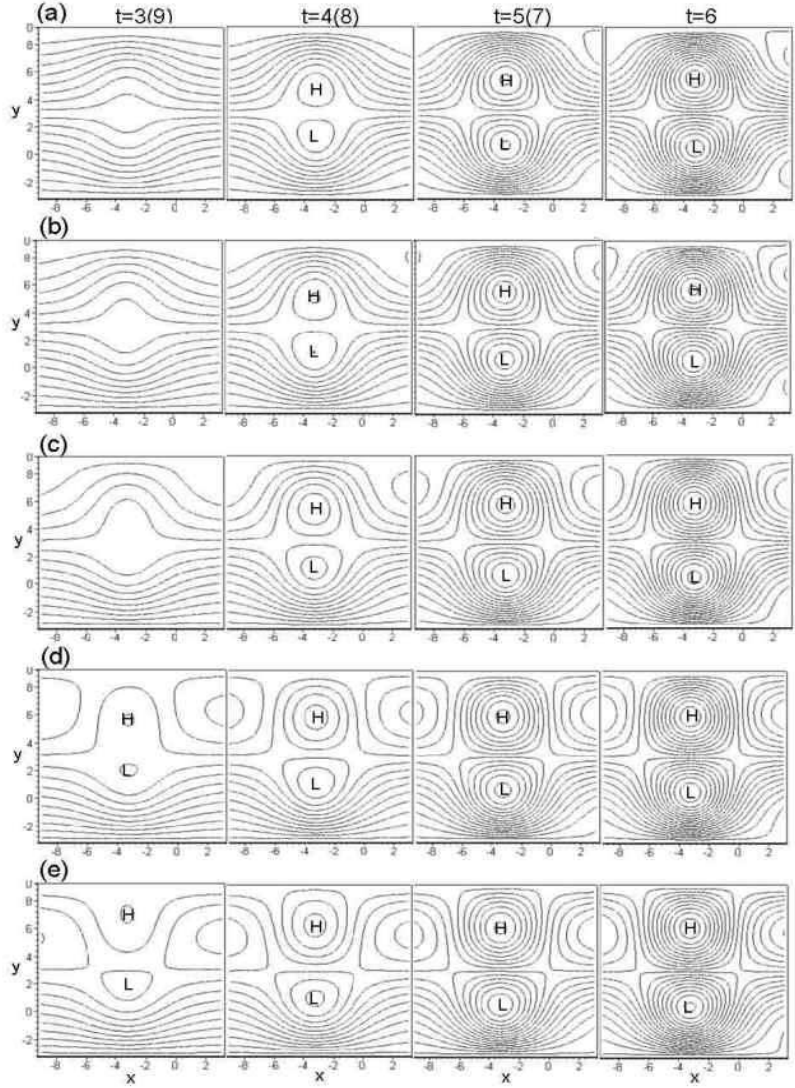}
\centering\caption{The stream function patterns of DB evolution
under different anticyclonic westerly shears. (a) $a_2=-0.03$; (b)
$a_2=-0.15$; (c) $a_2=-0.3$; (d) $a_2=-0.45$; (e) $a_2=-0.6$.
CI=0.4.}
\end{figure}

\subsection{Effect of the time-dependent background westerlies}

In the real atmospheric circulation, the background westerly is
not always a constant during the life period of blocking. Actually
there is interaction between blocking and the background
westerlies; that is, the weak background westerlies modulated by
transient eddies are a precondition for the onset of blocking;
meanwhile, after the blocking is established, it prevents the
westerlies passing through as a result of a weakened westerly
(Berggren et al. 1949; Elliott and Smith 1949; Egger et al. 1986;
Mullen 1987; Long 1964; Shutts 1983; Colucci 1985; Holopainen and
Fortelius 1987; Dole 1989; Luo et al. 2001). Therefore, the
background westerlies may be a time-dependent function during the
episode interacting with blocking. For simplification, the
westerly shear ($a_2=0$) is not considered here, so the background
flow, $\overline{u}=\overline{u}_0-\epsilon a_3(t)$, here is
$\overline{u}_0=-a_1=0.5$. Several types of $a_3(t)$ profiles
shown in Fig. 9 are attempted to investigate the influence of
time-dependent background westerlies (TDW) on DB evolution. Case 1
indicates the condition that the background westerlies become
weakest at DB onset stage and increases gradually throughout the
block life cycle. It is found that the variations of the curve
slope do not impact the evolution of DB in Case 1. Case 2 and 3
show the westerly profiles reflecting the interaction between the
background westerlies and blocking with the westerlies decreasing
during the developing period before the DB peak at $t=6$ and
increasing throughout the decay stage after the DB peak. The
latter is conceptually consistent with the actual variation of the
westerlies during a blocking evolution (Elliott and Smith 1949).

\input epsf
\begin{figure}
     \includegraphics[bb=250 300 300 500,scale=0.9]{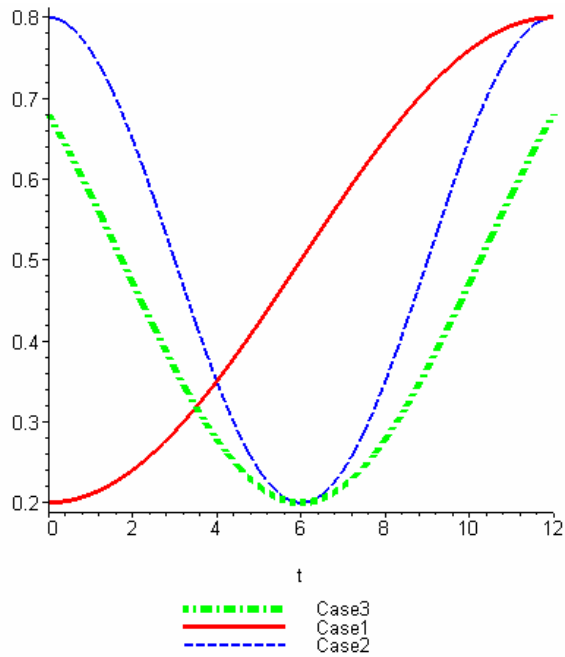}
     \centering\caption{Different types of TBW
     profiles during the DB life cycle. The x-axis is
     time $t$ in DB life episode and y-axis denotes
     $\overline{u}$. Case1 represents $\overline{u}=0.5-0.3cos(0.26t)$,
     Case2 denotes $\overline{u}=0.5-0.3cos(0.52(t-6))$,
     Case3 denotes $\overline{u}=0.5-0.3cos(0.37(t-6))$.}
\end{figure}

Fig. 10 presents the intensity of the DB high (Fig. 10a) and low
(Fig. 10b) center varying with respect to time $t$ under different
types of TDW as shown in Fig. 9 during the DB life cycle. For case
1, the TDW varying from the weakest to the strongest stage during
the DB life cycle, shortens the life period of the DB and weakens
the intensity of the DB. Tests changing slopes of the
($\overline{u}-t$) profile indicate that variation rate of the
background westerlies during the DB episode does not act as an
influence on the DB evolution (Figures omitted). This result
implies that the time-dependent westerlies varying accordantly
before and after the DB reaching its peak phase play a similar
role in the DB evolution. For the TDW reflecting interaction
between blocking and background flow in case 2, the evolution of
DB has been considerably impacted. The most noticeable feature is
that the time of the DB onset and decay lags that without TDW
term, either for the high (Fig. 10a) or for the low center (Fig.
10b) in the dipole structure. The decay process from the peak
phase of DB to its disappearance is longer than the DB developing
process from the onset of DB to its peak phase, resulting in the
asymmetric evolution of the DB during its life episode. In
addition, the intensity of the DB high/low centers at its peak
phase strengthens slightly. However, the intensity of DB depends
on the slope of the profile. For example, when the slope is not so
sharp (case 3 in Fig. 9) as in case 2, the high (low) peak drops
down (goes up) indicating the weakening of the DB (dot-dashed line
in Fig. 10), but the lag and asymmetric characteristics are not
changed.

\input epsf
\begin{figure}
     \includegraphics[bb=100 300 500 600,scale=0.9]{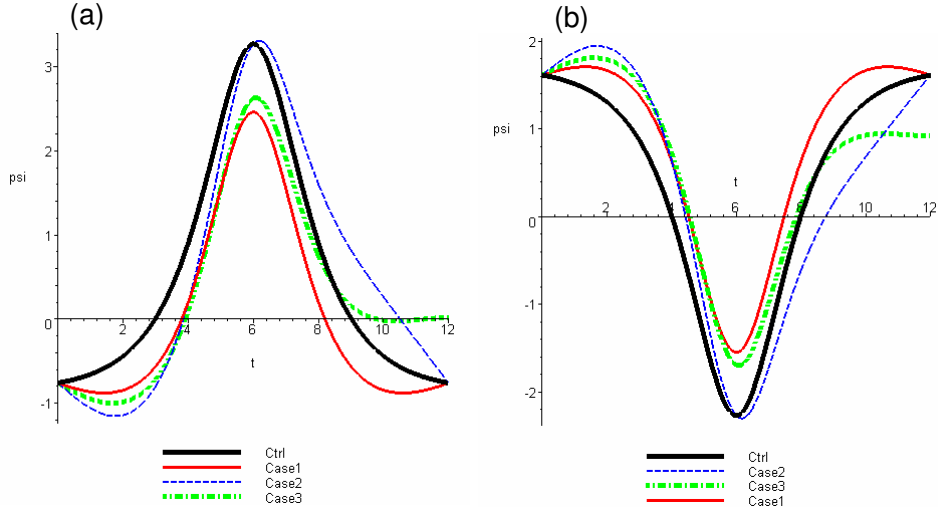}
     \centering\caption{The intensity of the DB high (a) and low (b) center
     varying with respect to time $t$ under different types of
TBW as shown in Fig. 9 during the DB life cycle. Label "Ctrl"
means the case of background westerly $\overline{u}$ without
time-dependent term $a_3(t)=0$.}
\end{figure}

The detailed evolutions of DB impacted by TDW are presented in
Fig. 11. For case 1 (Fig. 11a), the DB evolution process is
similar to that without the TDW term in Fig. 3b, exhibiting a
symmetric feature about the peak phase in the evolution of the DB
episode. The most differences between Fig. 3b and Fig. 11a are in
the different intensity of DB at any phase during the DB life
cycle and zonal and meridional scales. The high/low intensity
weakens in case 1 and the zonal and meridional scales of DB all
lessens a little, suggesting the DB period shortens, which is
coherent with the results discussed above from Fig. 10. For case 2
the developing process from the onset ($t=3$) of DB to its peak
phase ($t=6$) resembles that in case 1, but the decay process from
the DB peak phase to blocking vanishing ($t=10$) appears very
different. Compared to the DB evolution process without TDW (Fig.
3b), the dipole center of the blocking in case 2 is stronger than
that in Fig. 3b at the beginning decay stage ($t=7, 8$). At $t=9$
the dipole-like pattern with closed high/low centers still
maintains although the streamlines become sparser in case 2, while
it disappears and only shows a dipole-like envelop in Fig. 3b. At
$t=10$ the almost straight streamlines in Fig. 3b are instead of
dipole-like envelop in Fig. 11b with sparser streamlines. Above
all, the effect of the TDW on evolution of DB is owing to altering
the DB life period and leading to the asymmetry of the DB life
cycle evolution with respect to its peak phase.

\begin{figure}
\includegraphics[bb=70 300 800 700,scale=0.75]{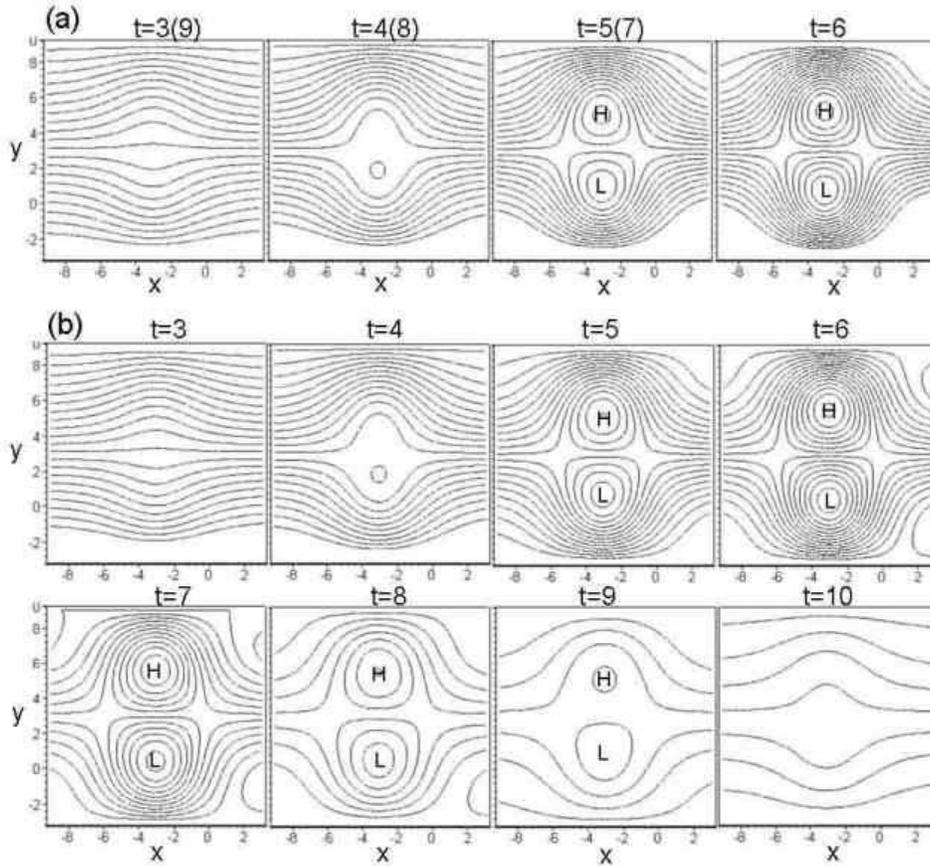}
\centering\caption{The stream function patterns of DB evolution
under different time-dependent westerly profiles appeared in Fig.
9. (a) Case 1; (b) Case 2. CI=0.4.}
\end{figure}

\section{Observational features of DB in the Northern Hemisphere}

\subsection{Data and the DB definition}

The data source in this study is the NCEP-NCAR reanalysis data
from January 1958 to December 1997, which uses a state-of-the-art
global data assimilation system (Kalnay et al. 1996). The
variables used in this study are daily mean geopotential height at
1000 and 500 hPa and the west-east wind component at 500 hPa, on a
$2.5^\circ$ latitude $\times 2.5^\circ$ longitude grid, from
$20^\circ N$ to $90^\circ N$ around the Northern Hemisphere.

Although the blocking phenomenon is well known in the
meteorological community, there is no generally accepted
definition of a blocking event (Lejen\"and $\varnothing$kland
1983). Commonly used definitions can be divided into four
categories of methods to identify blocking. The first is a
subjective technique first put forward by Rex (1950a,b) in the
1950s. This was then inherited and modified by Sumner (1954) and
White and Clark (1975). They identified the blocking events
subjectively by visual inspection and then used semiobjective
criteria to determine the exact dates of initiation and duration
of blocking based on examination of the daily weather charts. The
second category uses objective criteria first used by Elliott and
Smith (1949) based on magnitude and persistence of pressure
anomalies. Later, Hartman and Ghan (1980), Dole and Gordon (1983),
Dole (1986, 1989), Shukla and Mo (1983), Huang (2002), and Huang
et al. (2002b) extended the analysis from sea level pressure
departure to the occurrence of persistent positive geopotential
height anomalies at the upper level. The useful objective criteria
for identifying blocking events were designed by Lejen\"and
$\varnothing$kland 1983, using the north-south geopotential height
gradient based on the coherence between the occurrence of
persistent anomalous mid-latitude easterly flow and blocking. The
advantage of this objective method is its simplicity for automatic
calculation, and it has therefore been used widely (Tibaldi and
Molteni 1990; D＊Andrea et al. 1998; Huang et al. 2002a).
Recently, Pelly and Hoskins (2003a,b) constructed a new dynamical
blocking index using a meridional $\theta$ difference on a
potential vorticity (PV) surface. They reveal that their
PV-$\theta$ index is better able to detect $\Omega$ blocking than
conventional height field indices. Since we are focusing on
dipole-type blocking, the newest PV-$\theta$ index is not suitable
for us in this study.

All of the definitions of blocking mentioned above do not wholly
address DB, so new criteria for identifying DB have to be
established. In this study the DB is distinguished by the
following criteria.

1) At least on pair of closed high/low contours appears
simultaneously at 4 (2.5) geopotential decameters (dam) contour
interval at 500 hPa (1000 hPa), with the distance between the high
and low centers is no larger than 30 longitudes.

2) The westerlies must split into two branches at 500 hPa, and the
distance between the divarication point and the meeting point is
no less than 45 longitudes.

3) The closed high or low center lasts at least 5 days.

4) The moving speed of the blocking does not exceed ten degrees of
longitude per day.

5) The high center is located at least north of $40^{\circ} N$.

6) The blocking index (BI) is no less than 20 $m s^{-1}$.

The BI is calculated as $BI=U_{WM}-3\times U_{EM}$ $m s^{-1}$,
where $U_{EM}$ is the maximum of the 20-point running mean of
geostrophic easterlies bounded by 20 degrees of longitude in the
east and west of the high center, respectively, and 30 degrees of
latitude south of the high center; $U_{WM}$ is the maximum of the
westerlies south of the $U_{EM}$ position bounded by 30 degrees of
latitude. Because the easterlies south of the blocking anticyclone
are usually about 3 times less than the westerlies, 3 times
$U_{EM}$ is calculated in BI. The geostrophic wind $U$ is
calculated as $U=-(g/f)\partial z/\partial y$ ($f$ is the Coriolis
parameter, $g$ is the gravitational acceleration, and $z$ is the
geopotential height at 500 hPa). The latitude and longitude of the
high center denote the position of the high, which is prescribed
similarly to that of Dole (1986) as persistent positive anomalies
(with respect to latitude mean) greater than 70 gpm at 500 hPa.
The number of times that dipole-type blocking appears is counted
and the latitude where the high center of a dipole block is
located is regarded as the meridional position of the DB.

\subsection{Climatological features of DB and possible reasons}

Many studies have derived a comprehensive set of climatological
statistical characteristics of blocking anticyclones using
subjective or objective techniques, including location, frequency,
duration, intensity, size, and distribution (Elliott and Smith
1949; Rex 1950a,b; White and Clark 1975; Lupo and Smith 1995a;
Lejen\"and $\varnothing$kland 1983; Huang et al. 2002a; Huang
2002). Preferred Northern Hemisphere locations for blocking are
the northeastern boundaries of the Pacific and Atlantic Oceans and
the Ural area (Elliott and Smith 1949; Rex 1950b; Lejen\"and
$\varnothing$kland 1983; Dole and Gordon 1983; Shukla and Mo 1983;
Huang 2002). For the statistical characteristics of DB, only Luo
and Ji (1991) performed an observational study of dipole-type
blocking in the atmosphere during 1969-84. However, the time
series are too short to include the interdecadal variability of
blocking (Huang et al. 2002b). The availability of the NCEP-NCAR
reanalysis data (Kalnay et al. 1996) includes a considerably long
time series, and so provides a solid basis for understanding
behavior of the atmospheric midlatitude blocking.

In this paper, statistical features of DB in the Northern
Hemisphere are investigated based on a 40-yr period data spanning
from 1958每97. Results show that there are three preferable
regions: the northwestern Pacific, the Atlantic, and the Ural
area, where DB frequently occurred, agreement with the results of
Luo and Ji (1991). The Pacific is the most preferable region,
where a total of 1982 days of DB occurred, far more than the
Atlantic (848 days) and Ural areas (533 days). It is interesting
to note that the papers by Elliott and Smith (1949) and Rex
(1950b), which were published within a year of each other, gave
contradictory results on the relative frequency of blocking in the
Atlantic and Pacific. White and Clark (1975), however, obtained a
result that was not in agreement with Rex (1950b). In the research
of Elliott and Smith (1949), they found that the total number of
blocking days in central Pacific is 4 times of that in the
northeastern Atlantic. Notice that the previous results did not
distinguish the DB or monopole blocking, Elliott and Smith＊s
(1949) results might include more DBs, while the statistics of Rex
(1950b) may include more monopole blocks since they used different
data in different periods of time or their definitions of blocking
may not be exactly the same.

The latitudinal distribution of DB over the three favorable
regions is displayed in Fig. 12. It is shown that the latitude of
DB over the Pacific is mostly concentrated at $65^\circ N\sim
80^\circ N$, while the DB over Atlantic focuses at $55^\circ N\sim
65^\circ N$, about $10^\circ$ southward from that in the Pacific.
The Ural area has the least DB, mainly at $60^\circ N\sim 70^\circ
N$, and it is also more southward than that over the Pacific.

\begin{figure}
     \centering\epsfxsize=10cm\epsfysize=6cm\epsfbox{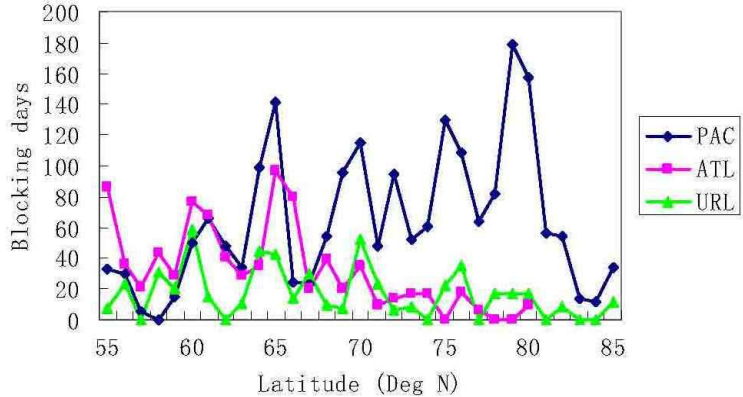}
     \caption{The latitudinal distribution of DB over three favorable regions.}
\end{figure}

From the statistics above it is found that the Pacific is the most
favorable place for DB occurrence, but fewer blocking events
(include DB and $\Omega$ blocking) occurred over the Pacific
compared to the Atlantic and Ural areas (see Huang 2002). Why DB
seems to prefer the Pacific? This is really a puzzling question.
According to the analysis in section {\ref{theory}, we know the
intensity of the background mean westerlies is associated with the
DB development. Therefore, the observational westerlies are
examined first. Figure 13 displays the climatological seasonal
cycle of the westerlies averaged in the high latitudes
($55^\circ-80^\circ N$) for the Northern Hemispheric DB preferable
latitudes at 500 hPa and annual mean westerly profiles with
respect to latitudes averaged over three favorable regions for DB
occurrence: the Pacific ($120^\circ E-150^\circ W$), the Atlantic
($60^\circ W-30^\circ E$) and the Ural ($30^\circ E-90^\circ E$)
areas, respectively. It is noticeable that the westerlies where
the high center of DB locates preferably over the high latitudes
of the Pacific region are almost all less than that over the
Atlantic and Ural areas (Fig. 13a). The climatological annual mean
westerly over the Pacific is about $3.25 m/s$, not reaching half
of that over the Atlantic ($7.35 m/s$) or Ural areas ($7.77 m/s$).
Hence it is not strange that many more DBs occurred over the
Pacific than over the Atlantic and Ural regions. From a
climatological point of view, the weak westerlies are one of the
most important factors that impact the DB occurrence.

However, the weak westerly is a precondition for onset of
blocking, including DB and monopole blocks (Shutts 1983, 1986; Luo
and Ji 1991; Luo 1994; Luo et al. 2001). The inconsistency between
DB (the most located at northwestern Pacific) and total blockings
(the least occurred at northeastern Pacific) implies that the weak
westerly is a necessary condition for blocks, but not a sufficient
condition for DB onset. Since the westerly shears play a
considerably important role in DB evolution, the annual mean
westerly profiles with respect to latitudes averaged over the
three DB preferable regions are also studied (Fig. 13b). The
figure reveals that westerly profile over the global DB regions
appears appears to be a weak cyclonic westerly shear from
$40^\circ N$ to $80^\circ N$, which is advantageous for DB
establishment at these latitudes. The Pacific has the strongest
CWS from $40^\circ N$ to $60^\circ N$ and a considerably weak AWS
from $60^\circ N$ to $80^\circ N$, which favors more DB produced
at higher latitudes and less DB reduced at lower latitudes,
corresponding to the latitudinal distribution of the DB shown in
Fig. 12.

\input epsf
\begin{figure}
     \centering\epsfxsize=13cm\epsfysize=7cm\epsfbox{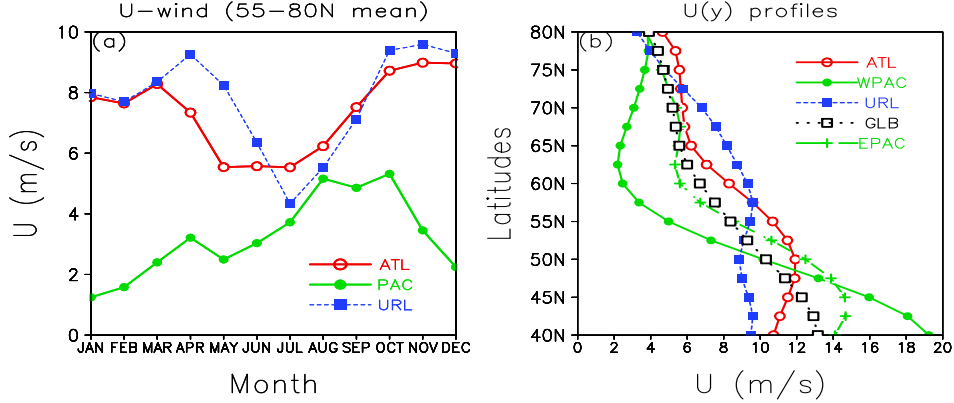}
     \caption{Climatological seasonal cycle of westerlies (a) and
     westerly profiles with respect to latitudes (b) averaged over
     Pacific (PAC or WPAC, $120^\circ E\sim 150^\circ W$),
     Atlantic (ATL, $60^\circ W\sim 30^\circ E$) and
     Ural (URL, $30^\circ E\sim 90^\circ E$) areas. Label "EPAC" denotes
     the eastern Pacific averaged from $180^\circ$ to $90^\circ W$,
     and "GLB" is the mean of PAC, ATL and URL.}
\end{figure}

Observational studies also give the seasonal variability of global
DB as shown in Fig. 14a. The numbers of DB decrease from winter
(December to February) to spring (March to May), summer (June to
August), and autumn (September to November) successively. Winter
is the most favorable season for DB producing, consistent with the
results for total blocking by White and Clark (1975), Lejen\"and
$varnothing$kland (1983), Shukla and Mo (1983), and Huang (2002),
but contradictory to results by Rex (1950a); while autumn is the
quiet season for DB onset, which disagrees with the common
statistical results for total blocking (White and Clark 1975;
Lejen\"and $varnothing$kland 1983; Shukla and Mo 1983; Huang
2002). The mean westerlies averaged over the three preferable DB
regions in each season (Fig. 14b) indicate that the autumn
westerlies are the strongest, and are associated with the least DB
occurring in autumn according to the theory discussed above. The
summer westerly is weakest, but does not correspond to the largest
DB frequency in summer. Nevertheless, the seasonal variability of
the background westerlies over the Pacific region ($120^\circ
E\sim 150^\circ W$) in Fig. 14c appears to have a completely
negative correlation to the seasonal cycle of the global DB
occurrence in Fig. 14a. This suggests that the seasonal variation
of the global DB is mostly due to the seasonal cycle of the
Pacific DB, which is mostly determined by the mean background
westerlies over the Pacific. For the Atlantic and Ural DB, the
seasonal variability of the mean westerlies may play important but
indecisive role in the DB seasonal cycle. The seasonal westerly
profiles with respect to latitudes over the Pacific region of DB
(Fig. 14d) show that the westerly shears are very favorable for DB
establishment in winter, spring and autumn, with a CWS at
$45^\circ-60^\circ N$ and very weak AWS at higher latitudes, while
in summer the AWS at $60^\circ-75^\circ N$ seems too strong to be
unfavorable for DB pattern developing. It is remarkable that among
the very weak AWSs in winter, spring and autumn, the AWS in winter
is the relatively strongest one, which is favorable for strong DB
development, and results in the most DB occurring in winter.
Therefore, the intensity of the background mean westerlies and
their shear structures closely associate with the establishment of
DB, leading to the northwestern Pacific being the most preferable
region for DB and its seasonal variability.

\input epsf
\begin{figure}
\centering\epsfxsize=12cm\epsfysize=9cm\epsfbox{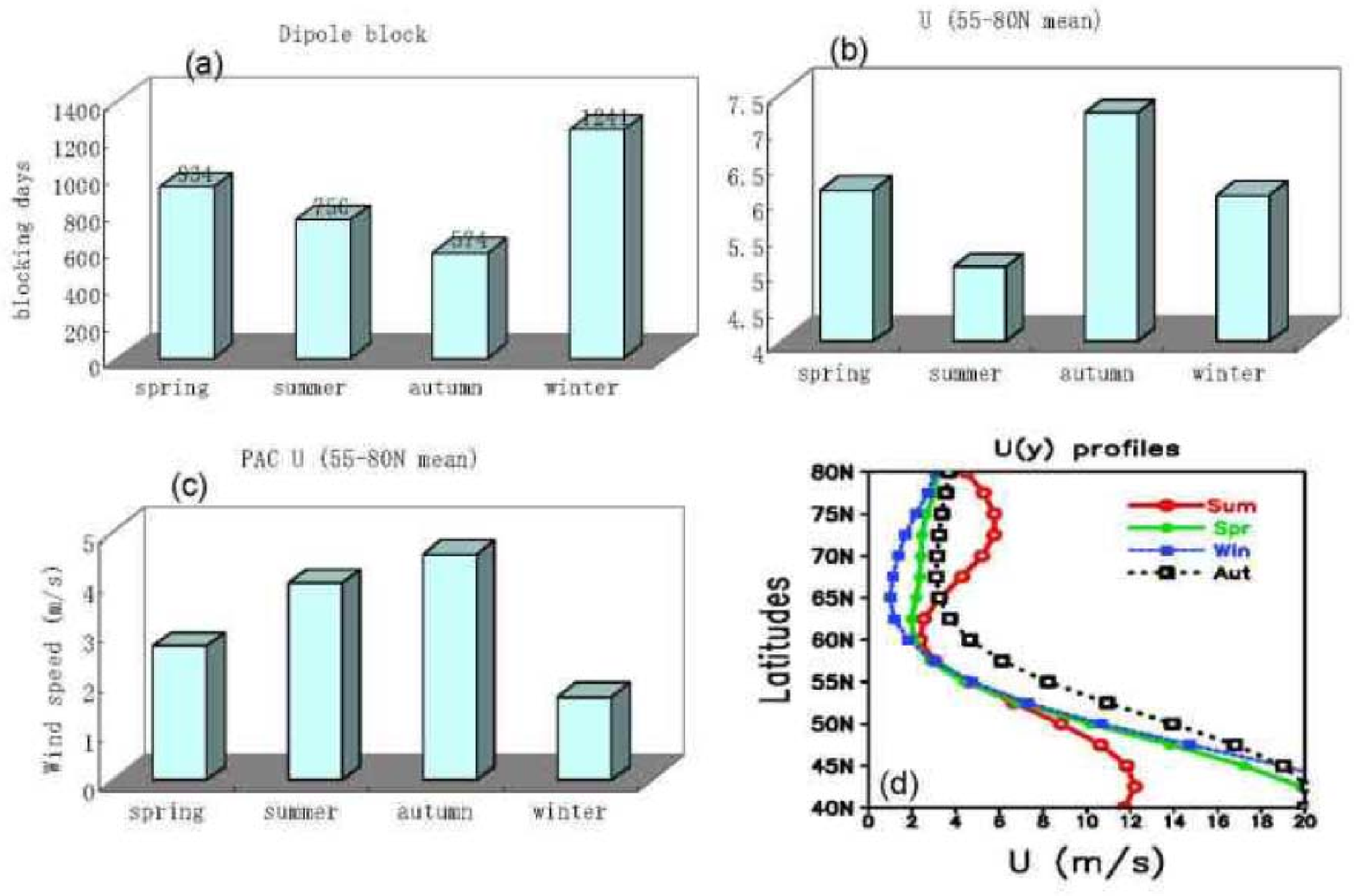}
     \caption{Seasonal variations of (a) the number of DB days, (b) the
     mean westerlies averaged over the three DB preferable regions, (c) westerlies averaged over
     the Pacific ($120^\circ E\sim 150^\circ W$), and (d) seasonal mean westerly profiles
     with respect to latitude averaged over
     the Pacific. Label "Win" denotes winter season (December to February),
     "Sum" denotes summer (June to August), "Spr" denotes spring (March to
     May),
      and "Aut" denotes autumn (September to November).}
\end{figure}

\section{Concluding remarks}

In this paper, a variable coefficient Korteweg de Vries (VCKdV)
system is derived by considering the time dependent background
flow and boundary conditions from the nonlinear, inviscid,
nondissipative, and equivalent barotropic vorticity equation in a
beta-plane channel. The analytical solution obtained from the
VCKdV equation can be successfully used to explain the evolution
of atmospheric dipole-type blocking life cycle. Analytical
diagnoses are analyzed and three factors that may influence the
evolution of atmospheric DB are investigated.

Theoretical results show that the background mean westerlies have
great influence on evolution of DB during its life episode. Weak
westerlies are necessary for blocking development and the high/low
centers of blocking decrease and move southward and westward, the
horizontal scale enlarges and meridional span reduces, as well as
the blocking life period shortens with respect to the enhanced
westerly.

Shear of the background westerlies also plays an important role in
evolution of DB. The CWS is preferable for the development of DB,
which agrees with the recent result from Luo (2005b). When the
cyclonic shear increases, the intensity of DB decreases and its
life period becomes shorter. Weak AWS is also favorable for DB
formation, but a critical threshold shear exists, beyond which a
cutoff anticyclone from the North Pole region develops
dramatically instead of an envelope Rossby soliton forming the
high anticyclone center of DB. Inside the critical threshold of
the anticyclonic shear, the intensity of DB increases and the life
period of DB prolongs when the anticyclonic shear increases. For
the role of CWS in DB evolution, our result is inconsistent with
that of Luo (2005b), who revealed that an isolated vortex pair
block excited resonantly by synoptic-scale eddies is more likely
suppressed in an anticyclonic shear environment. The difference
may lie on different nonlinear systems considered by Luo (2005b)
and us. That is, Luo (2005b) considered an eddy-forced nonlinear
Schr\"{o}dinger Rossby soliton system in the atmosphere, while we
established a nonlinear system based on a nonlinear KdV equation
without forcing, and our results show that the nonlinear effect of
the free atmosphere could also induce blocking circulation by its
intrinsic nonlinear interaction without external eddy forcing.

Time-dependent variation of background flow in the life cycle of
DB has some modulations on blocking life period and intensity due
to the behavior of the mean westerlies. The effect of TDW,
especially the style of TDW, reflecting interaction between the
background westerlies and blocking, with the westerlies decreasing
during the developing period before DB peak and increasing
throughout the decay stage after DB peak, is caused by the
alteration of the DB life period and leads to the asymmetry of the
DB life cycle evolution.

Statistical analysis of climatological features of observed DB is
investigated based on 40-yr geopotential height fields from the
NCEP-NCAR reanalysis data during 1958-97. Observational results
show that there are three preferable regions of DB in the Northern
Hemisphere, located in the northwestern Pacific, the northeastern
Atlantic and the Ural Mountain areas, in agreement with the
results of Luo and Ji (1991). The number of DB days occuring over
the Pacific is far larger than total of that over the other two
regions, corresponding to the weaker westerlies over the
northwestern Pacific and the particular westerly shear structure
over Pacific. The Pacific DB prefers to be established at higher
latitudes than the Atlantic and Ural regions by about $10^\circ$,
which may be caused by its having the strongest CWS at
mid-latitudes and weaker AWS at higher latitudes over the Pacific
region. Seasonal variability of the global DB is also associated
with the seasonal cycle of the mean westerlies over the Pacific
and the westerly shear structure. Therefore, the intensity of the
mean westerlies and the shear structure of the westerly profiles
are two important conditions associated with the climatological
features of the DB in the Northern Hemisphere, which may play
crucial role in the DB life cycles. The type of TDW could also
impact the DB life episode concluded from analytical resolution,
but it is yet to be demonstrated from the observational studies in
the future.

\begin{acknowledgements}
The work was supported by the National Natural Science Foundation
of China (Grants 40305009, 10475055, 1054124, and 90203001),
Program for New Century Excellent Talents in University
(NCET-05-0591), Program 973 (No. 2005CB422301), and Shanghai
Post-Doctoral Foundation.
\end{acknowledgements}

\centerline{\bf REFERENCES}

 Berggren, R., B. Bolin, and C. G. Rossby, 1949: A
aerological study of zonal motion, its perturbations and
breakdown. Tellus, 1, 14每37.

Butchart, N., K. Haines, and J. C. Marshall, 1989: A theoretical
and diagnostic study of solitary waves and atmospheric blocking.
J. Atmos. Sci., 46, 2063每2078.

Cascaval, R. C., 2003: Variable coefficient korteveg-de vries
equations and wave propagation in elastic tubes. Evolution
Equations, Goldstein, G. and Nagel, R. and Romanelli, S. (eds.),
New York, NY: Marcel Dekker, 57每69.

Charney, J. G. and J. G. DeVore, 1979: Multiple flow equilibria in
the atmosphere and blocking. J. Atmos. Sci., 36, 1205每1216.

Chen, W. and H. H. Juang, 1992: Effect of transient on blocking
flows: General circulation model experiments. Mon. Wea. Rev., 120,
787每801.

Colucci, S. J., 1985: Explosive cyclogenesis and large-scale
circulation changes: Implications for atmospheric blocking. J.
Atmos. Sci., 42, 2701每2717.

-- 1987: Comparative diagnosis of blocking versus non-blocking
planetary-scale circulation changes during synoptic-scale
cyclogenesis. J. Atmos. Sci., 44, 124每139.

Colucci, S. J. and T. L. Alberta, 1996: Planetary-scale
climatology of explosive cyclogenesis and blocking. Mon. Wea.
Rev., 124, 2509每2520.

Colucci, S. J. and D. P. Baumhefner, 1998: Numerical prediction of
the onset of blocking: a case study with forecast ensembles. Mon.
Wea. Rev., 126, 733每784.

D'Andrea, F., S. Tibaldi, and M. Blackburn, 1998: Northern
hemisphere atmospheric blocking as simulated by 15 atmospheric
general circulation models in the period 1979-1988. Climate
Dynamics, 14, 385每407.

Dole, R. M., 1986: Persistent anomalies of the extratropical
northern hemisphere wintertime circulation: Structure. Mon. Wea.
Rev, 114, 178每207.

-- 1989: Life cycle of persistent anomalies, part i: Evolution of
500-mb height fields. Mon. Wea. Rev, 117, 177每211.

Dole, R. M. and N. D. Gordon, 1983: Persistent anomalies of the
extra-tropical northern hemisphere wintertime circulation:
Geographical distribution and regional persistence
characteristics. Mon. Wea. Rev, 111, 1567每1586.

Egger, J., W. Metz, and G. M“uller, 1986: Synoptic-scale eddy
forcing of planetaryscale blocking anticyclones. Adv. Geophys.,
29, 183每198.

Elliott, R. D. and T. B. Smith, 1949: A study of the effects of
large blocking highs on the general circulation in the northern
hemisphere westerlies. J. Meteor., 6, 67每85.

Flytzanis, N., S. Pnevmatikos, and M. Remoissenet, 1985: Kink,
breather and asymmetric envelop or dark solitons in nonlinear
chains: I. monatomich chain. J. Phys. C, 18, 4603每4629.

Gottwald, G. A. and R. H. J. Grimshaw, 1999: The effect of
topography on dynamics of interacting solitary waves in the
context of atmospheric blocking. J. Atmos. Sci., 56, 3663每3678.

Haines, K. and A. J. Holland, 1998: Vacillation cycles and
blocking in channel. Quart. J. Roy. Meteoro. Soc., 124, 873每895.

Haines, K. and J. C. Marshall, 1987: Eddy-forced coherent
structures as a prototype of atmospheric blocking. Quart. J. Roy.
Meteoro. Soc., 113, 681每704.

Hansen, A. and T. C. Chen, 1982: A spectral energetics analysis of
atmospheric blocking. Mon. Wea. Rev., 110, 1146每1159.

Hartman, D. L. and S. J. Ghan, 1980: A statistical study of the
dynamics of blocking. Mon. Wea. Rev., 108, 1144每1159.

Holopainen, E. and C. Fortelius, 1987: High-frequency transient
eddies and blocking. J. Atmos. Sci., 44, 1632每1645.

Hoskins, B. J., M. E. McIntyre, and A.W. Robertson, 1985: On the
use and significance of isentropic potential vorticity maps.
Quart. J. Roy. Meteoro. Soc., 111, 877每946.

Huang, F., 2002: Study on the atmospheric blocking circulation
over the North Pacific during winter and its connection with the
mid-low latitude interaction. Ph.D. thesis, Ocean University of
China.

Huang, F., F. X. Zhou, and P. J. Olson, 2002a: Variations of the
atlantic and pacific blocking anticyclones and their correlation
in the northern hemisphere. J. of Ocean University of Qingdao
(English Edition), 1, 38每44.

Huang, F., F. X. Zhou, and X. D. Qian, 2002b: Interannual and
decadal variability of north pacific blocking and its relationship
to sst, teleconnection and storm track. Adv. Atmos. Sci., 19,
807每820.

Illari, L., 1984: A diagnostic study of the potential vorticity in
a warm blocking anticyclone. J. Atmos. Sci., 41, 3518每3526.

Illari, L. and J. C. Marshall, 1983: On the interpretation of eddy
fluxes during a blocking episode. J. Atmos. Sci., 40, 2232每2242.

Ji, L. R. and L. Tibaldi, 1983: Numerical simulations of a case of
blocking: The effects of orography and land-sea contrast. Mon.
Wea. Rev., 111, 2068-2086.

Kalnay, E., M. Kanamitsu, R. Kistler, W. Collins, D. Deaven, L.
Gandin, M. Iredell, S. Saha, G. White, J. Woollen, Y. Zhu, M.
Chelliah, W. Ebisuzaki, W. Higgins, J. Janowiak, K. C. Mo, C.
Ropelewski, J.Wang, A. Leetmaa, R. Reynolds, R. Jenne, and D.
Joseph, 1996: The ncep/ncar 40-year reanalysis project. Bull. Am.
Meteorol. Soc., 77, 437-471.

Lejen\"as, H. and H. $\varnothing${land}, 1983: Characteristics of
northern hemisphere blocking as determined from a long time series
of observational data. Tellus, 35A, 350-362.

Long, R. R., 1964: Solitary waves in the westerlies. J. Atmos.
Sci., 21, 197每200.

Luo, D. H., 1994: The dynamical characters of local blocking high
and dipole blocking in the atmosphere. Plateau Meteorology (in
Chinese), 13, 1每13.

-- 1995: Solitary rossby waves with the beta parameter and dipole
blocking. Quarterly Journal of Applied Meteorology (in Chinese),
6, 220每227.

-- 1999: Envelope soliton theory and blocking pattern in the
atmosphere. China Meteorological Press, 113 pp.

-- 2000: Nonlinear dynamics of blocking. China Meteorological
Press, 248 pp.

-- 2001: Derivation of a higher order nonlinear schrodinger
equation for weakly nonlinear rossby waves. Wave Motion, 33,
339每347.

-- 2005a: A barotropic envelope rossby soliton model for
block-eddy interaction. part ii role of westward-traveling
planetary waves. J. Atmos. Sci., 62, 22每40.

-- 2005b: A barotropic envelope rossby soliton model for
block-eddy interaction. part iv: Block activity and its linkage
with sheared environment. J. Atmos. Sci., 62, 3860每 3884.

Luo, D. H., F. Huang, and Y. N. Diao, 2001: Interaction between
antecedent planetaryscale envelope soliton blocking anticyclone
and synoptic-scale eddies: Observations and theory. J. Geophy.
Res., 106, 31795每31815.

Luo, D. H. and L. R. Ji, 1991: Observational study of dipole
blocking in the atmosphere. Scientia Atmospherica Sinica (in
Chinese), 15, 52每57.

Luo, D. H., J. P. Li, and F. Huang, 2002: Life cycles of blocking
flows associated with synoptic-scale eddies: observed results and
numerical experiments. Adv. Atmos. Sci., 19, 594每618.

Luo, D. H. and H. Xu, 2002: The dynamical characters of local
blocking high and dipole blocking in the atmosphere. Journal of
Ocean University of Qingdao (in Chinese), 32, 501每510.

Lupo, A. R., 1997: A diagnosis of blocking events that occurred
simultaneously in the midlatitude northern atmosphere. Mon. Wea.
Rev., 125, 1801每1823.

Lupo, A. R. and P. J. Smith, 1995a: Climatological feature of
blocking anticyclones in the north hemisphere. Tellus, 47A,
439每456.

-- 1995b: Planetary and synoptic-scale interactions during the
life cycle of a midlatitude blocking anticyclone over the north
atlantic. Tellus, 47A, 575每596.

-- 1998: The interactions between a midlatitude blocking
anticyclone and synopticscale cyclone that occurred during the
summer season. Mon. Wea. Rev., 126, 502每 515.

Malguzzi, P. and P. Malanotte-Rizzoli, 1984: Nonlinear stationary
rossby waves on nonuniform zonal winds and atmospheric blocking,
part i: The analytical theory. J. Atmos. Sci., 41, 2620每2628.

McWilliams, J. C., 1980: An application of equivalent modons to
atmospheric blocking. Dyn. Atmos. Oceans, 5, 43每66.

Michelangeli, P. A. and R. Vautard, 1998: The dynamics of
euro-atlantic blocking onsets. Quart. J. Roy. Meteoro. Soc., 124,
1045每1070.

Mullen, S. L., 1987: Transient eddy forcing of blocking flows. J.
Atmos. Sci., 44, 3每22.

Pedlosky, J., 1979: Geophysical Fluid Dynamics. Springer-Verlag,
624 pp.

Quiroz, R. S., 1984: The climate of the 1983-84 winter: A season
of strong blocking and severe cold in north america. Mon. Wea.
Rev., 112, 1894每1912.

Rao, N. N., P. K. Shukla, and M. Y. Yu, 1990: Dust-acoustic waves
in dusty plasmas. Planet. Space Sci., 38, 543每546.

Rex, D. F., 1950a: Blocking action in the middle troposphere and
its effects upon regional climate: I. an aerological study of
blocking action. Tellus, 2, 196每211.

-- 1950b: Blocking action in the middle troposphere and its
effects upon regional climate: Ii. the climatology of blocking
action. Tellus, 2, 275每301.

Shukla, J. and K. C. Mo, 1983: Seasonal and geographical variation
of blocking. Mon. Wea. Rev., 111, 388每402.

Shutts, G. J., 1983: The propagation of eddies in diffluent
jetstreams: eddy vorticity forcing of blocking flow fields. Quart.
J. R. Meteor. Soc., 109, 737每761.

-- 1986: A case study of eddy forcing during an atlantic blocking
episode. Adv. Geophys., 29, 135每161.

Sumner, E. J., 1954: A study of blocking in the atlantic-european
sector of the northern hemisphere. Quart. J. R. Meteor. Soc., 80,
402每416.

Tanaka, H. L., 1991: A numerical simulation of amplification of
low frequency planetary waves and blocking formations by the
upscale energy cascade. Mon. Wea. Rev., 119, 2919每2935.

-- 1998: Numerical simulation of a life-cycle of atmospheric
blocking and the analysis of potential vortisity using a simple
barotropic model. J. Meteor. Soc. Japan, 76, 983每1008.

Tibaldi, S. and F. Molteni, 1990: On the opperational
predictability of blocking. Tellus, 42A, 343每365.

Tracton, M. S., K. Mo,W. Chen, E. Kalnay, R. Kistler, and G.
White, 1989: Dynamical extended range forecasting (derf) at the
national meteorological center. Mon. Wea. Rev., 117, 1604每1635.

Tsou, C. S. and P. J. Smith, 1990: The role of
synoptic/planetary-scale interactions during the development of a
blocking anticyclone. Tellus, 42A, 174每193.

Tung, K. K. and R. S. Linzen, 1979: A theory of stationary long
waves, part i:a simple theory of blocking. Mon. Wea. Rev., 107,
714每734.

Vautard, R. and B. Legras, 1988: On the source of midlatitude
low-frequency variability. part ii: Nonlinear equilibration of
weather regimes. J. Atmos. Sci., 45, 2845每 2867.

Vautard, R., B. Legras, and M. Deque, 1988: On the source of
midlatitude lowfrequency variability. part i: A statistical
approach to persistence. J. Atmos. Sci., 45, 2811每2844.

White, E. B. and N. E. Clark, 1975: On the development of blocking
ridge activity over the central north pacific. J. Atmos. Sci., 32,
489每501.

Yoshinaga, T. and T. Kakutani, 1984: Second order k-dv soliton on
the nonlinear transmission line. J. Phys. Soc. Jpn., 53, 85每92.

Zabusky, N. J. and M. D. Kruskal, 1965: Interaction of
§solitons§ in a collisionless plasma and the recurrence of
initial states. Phys. Rev. Lett., 15, 240每243.

\end{document}